\documentclass[twocolumn,preprintnumbers,endnote,nofootinbib,amsmath]{revtex4}
\usepackage{graphicx}
\usepackage{mathtools}
\usepackage[hypertex]{hyperref}
\newcommand{\vev}[1]{\langle {#1} \rangle}
\newcommand{\lsim}{\lesssim}
\newcommand{\gsim}{\gtrsim}

\newcommand{\beq}{\begin{equation}}
\newcommand{\eeq}{\end{equation}}
\newcommand{\eps}{\varepsilon}

\newcommand{\Gh}{SU(2)_h \times U(1)_h}
\newcommand{\mwh}{M_{W_h}}
\newcommand{\mzh}{M_{Z_h}}
\newcommand{\mah}{M_{\gamma_h}}
\newcommand{\ch}{\cos\theta_h}
\newcommand{\sh}{\sin\theta_h}
\newcommand{\sigannv}{\sigma_{\rm ann} v_{\rm rel}}
\newcommand{\sigel}{\sigma_{\rm el}}
\newcommand{\gah}{{\gamma_h}}

\begin{document}

\pagestyle{plain}

\title{\boldmath Dark Matter from Hidden Forces}

\author{Hooman Davoudiasl
}
\affiliation{Department of Physics, Brookhaven National Laboratory,
Upton, NY 11973, USA}

\author{Ian M. Lewis
}
\affiliation{Department of Physics, Brookhaven National Laboratory,
Upton, NY 11973, USA}


\begin{abstract}

We examine the possibility that dark matter may be
the manifestation of dark forces of a hidden sector, {\it i.e.} ``Dark Force = Dark Matter."  As an illustrative and minimal example we consider the hidden $\Gh$ gauge group.  The hidden dynamics is indirectly
coupled to the Standard Model (SM) through
kinetic mixing of $U(1)_h$ with the $U(1)_Y$ of hypercharge.  We assume a hidden
symmetry breaking pattern analogous to that of the SM electroweak symmetry, augmented with an
extra scalar that allows both the ``hidden $Z$ boson" $Z_h$ and the
``hidden photon" $\gamma_h$ to be massive.  The ``hidden $W$" bosons $W_h^\pm$
are dark matter in this scenario.  This setup
can readily accommodate a potential direct detection signal for dark matter at $\sim 10$~GeV
from CDMSII-Si data.  For some choices of parameters, the model can lead to
signals both in ``dark matter beam" experiments, from $Z_h\to W_h W_h$,
as well as in experiments that look for visible
signals of dark photons, mediated by $\gamma_h$.  Other possible phenomenological
consequences are also briefly discussed.

\end{abstract}
\maketitle

\section{Introduction}

The nature of dark matter (DM) remains a mystery.  Observational evidence
based on its gravitational effects suggests that DM
makes up roughly 25\% of the cosmic energy density \cite{Ade:2013zuv} and does not
have any significant non-gravitational interactions with ordinary matter.  This more or less sums up our
understanding of DM which is why there are a variety of currently viable
ideas for describing it.  So far, there is no clear signal of DM, but there are tentative hints
in various experiments, both astrophysical and terrestrial.  Some astrophysical data \cite{Adriani:2008zr} may be
accommodated by DM models that contain new light vector bosons that mediate dark sector forces with
feeble couplings to the visible Standard Model (SM) sector \cite{ArkaniHamed:2008qn}.  Of the laboratory signals, some hint at
a potential signal for light DM around $\sim 10$~GeV~\cite{Aalseth:2012if,Bernabei:2010mq,Angloher:2011uu,Petriello:2008jj}, the latest being from the CDMSII-Si data \cite{Agnese:2013rvf}.

Whether any of the aforementioned hints will grow in significance and rise to the level of a clear discovery
remains to be seen. Nonetheless, speculative thinking on this subject has led
to new models of DM, some of which postulate a dark sector endowed
with its own interactions, {\it i.e.} ``dark forces," characterized by scales near 1~GeV.
Many of these models use fermion or scalar DM candidates with an {\it ad-hoc} parity to guarantee stability.
However, if we use the SM as a guide, the stability of particles is expected to be derived from gauge symmetries,
Lorentz symmetry, or accidental symmetries that are a result of the gauge and Lorentz symmetries~\cite{parity}.

If the dark sector is endowed with a gauge symmetry, it may be possible to
assume that DM is made up of dark gauge fields whose mass has been generated by a Higgs mechanism.  This
possibility arises in cases where, after symmetry breaking, the non-abelian gauge bosons are
stable due to residual symmetries, obviating
the need for additional fermions or scalars.  Such minimal
scenarios - which imply ``Dark Force = Dark Matter" - have garnered
interest recently~\cite{VDMNonAbel,Chiang:2013kqa}.  However, these models
typically rely on a Higgs portal to communicate with the SM and
obtain the correct relic abundance\footnote{We note that Ref.~\cite{Chiang:2013kqa}
assumes a $U(1)_{B-L}$ gauge symmetry as part of the dark vector bosons to induce direct coupling to SM particles.}.
Here we point out that if we augment a non-abelian hidden gauge symmetry with an additional $U(1)_h$, the dark sector
and SM can interact via kinetic mixing between the $U(1)_h$ gauge boson and the SM hypercharge gauge boson.
Such a scenario can also lead to a rich phenomenology in low energy experiments~\cite{Bjorken:2009mm,
Reece:2009un,Andreas:2012mt,Bossi:2009uw,Merkel:2011ze,
Essig:2009nc,Babusci:2012cr,MeijerDrees:1992kd,Gninenko:2013sr,Adlarson:2013eza,
Abrahamyan:2011gv,Davoudiasl:2012ig,Endo:2012hp,Fayet:2007ua,Pospelov:2008zw,
Bennett:2006fi,Wojtsekhowski:2012zq,Boyce:2012ym,Davoudiasl:2012qa,darkZ,Gninenko:2012eq} if, as indicated by CDMSII-Si data,
the dark sector lives in the mass range $\lesssim \mathcal{O}(10)$~GeV.
For additional proposals with spin-1 DM candidates originating from other
sources see~Refs.~\cite{KKDM,Cheng:2003ju,VDMAbel}.

\section{The Model}
To show the viability of our scenario, we focus on the minimal higgsed gauge group with a non-abelian symmetry and kinetic mixing with the SM.  That is, we will assume a hidden gauge group $\Gh$ ({\it i.e.} a gauge group that the SM is uncharged under).
The hidden gauge symmetries are broken in close
analogy with the electroweak symmetry of the SM.
In particular, we will assume a hidden Higgs doublet $\Phi_h$ of $SU(2)_h$ with
charge 1/2 under $U(1)_h$,
leading to massive hidden vectors
$W_h^\pm$ by developing a non-zero vacuum
expectation value $\vev{\Phi_h}=(0,v_\Phi)^{\rm T}/\sqrt{2}$. In order to break the remaining $U(1)$, thereby giving mass to the ``hidden" photon, we
also introduce a complex scalar $\phi_h$ with
charge 1/2 only under $U(1)_h$ and $\vev{\phi_h}= v_\phi/\sqrt{2}$. In the following,
$g_h$ and $g_h'$ will denote the $SU(2)_h$ and $U(1)_h$
gauge couplings, respectively.

We have
\beq
W_{h\mu}^\pm = \frac{1}{\sqrt{2}} (W_{h\mu}^1 \pm i\,W_{h\mu}^2)\,,
\label{Wh}
\eeq
where $W_{h\mu}^i$, with $i=1,2,3$, are the $SU(2)_h$ gauge
fields and $\pm$ refers to ``hidden electric charges."  The masses of $W_h^\pm$ are given by the usual SM relation adapted to the hidden sector:
\beq
\mwh = \frac{g_h}{2}v_\Phi.
\eeq

The vev of the Higgs doublet, $v_\Phi$, leads to mixing between the other ``neutral" vector bosons (analogues of the SM $Z$ and photon)
parameterized by an angle $\theta_h$.  Denoting
these mass eigenstates by $Z_h$ and $\gamma_h$, we have
\beq
Z_{h\mu} = \cos \theta_h \, W_{h\mu}^3 - \sin \theta_h \, B_{h\mu}\,
\label{Zh}
\eeq
and
\beq
\gamma_{h\mu} = \sin \theta_h \, W_{h\mu}^3 + \cos \theta_h \, B_{h\mu},
\label{gamh}
\eeq
where $B_{h\mu}$ is the gauge field associated with $U(1)_h$.
One can show
\beq
\cos^2 \theta_h = \frac{\mwh^2 - \mah^2}{\mzh^2-\mah^2}\,,
\label{cos2thetah}
\eeq
where $\mwh$, $\mzh$, and $\mah$ denote the masses of $W_h$, $Z_h$, and $\gamma_h$, respectively.  We will
provide expressions for these masses in the Appendix.  Note that in the limit $\mah \to 0$ we recover the
SM-like relation ${\mwh^2 = \cos^2\theta_h \mzh^2}$ (with $\theta_h$
being the analogue of the weak mixing angle $\theta_W$).

While the SM fields do not carry charges under $\Gh$, the two sectors are assumed to be coupled through
a renormalizable kinetic mixing term \cite{Holdom:1985ag,darkphoton,nonabelian}
\beq
\frac{\eps}{2 \cos\theta_W} B_{h}^{\mu\nu} B_{\mu\nu}\,,
\label{kinmix}
\eeq
where $X_{\mu\nu} = \partial_\mu X_\nu - \partial_\nu X_\mu$, and $B_\mu$ is the SM hypercharge gauge field.  For simplicity we assume that kinetic mixing is the only portal between the dark sector and the SM, and ignore possible mixing in the Higgs sector.

Upon diagonalization of the kinetic terms in the usual way (see for example Ref.~\cite{darkZ} and the Appendix), one finds that
the two massive vectors, $\gamma_h$ and $Z_h$, couple to the visible electromagnetic current $J_{em}^\mu$
according to
\beq
{\cal L}_{\rm vh} = - \eps e \, [\ch  \,  \gamma_{h\mu} - \sh \,  Z_{h\mu}]J_{em}^\mu\,,
\label{EMcouplings}
\eeq
where $e$ is the SM electromagnetic coupling.
\section{Relic Density}
For $W_h$ to be viable as a dark matter candidate,  it needs to be cosmologically stable, have the correct
relic density $\Omega_{\rm DM}$, and not ruled out by direct or indirect searches.  With the above assumptions
about the hidden sector, the $W_h$ vectors are stable particles and will not decay.
This is due to a remnant $Z_2$ that persists after $\Gh$ breaking in this scenario\footnote{The effective operator $(\phi D_\mu \phi)^* (\Phi D^\mu\Phi)$ can mix $W_h^\pm$ with $B_h$, generating decays of $W_h^\pm$ to SM fermions.  To be a viable DM candidate, $W_h^\pm$ needs a lifetime of $\sim10^{27}$ seconds~\cite{Beringer:1900zz}.  With the assumptions in this paper, we find this operator needs to be suppressed by a mass scale $\gtrsim 10^{12}$ GeV.}.  To answer  the other questions,
we need to calculate the thermally averaged annihilation cross section $\vev{\sigannv}$,
with $v_{\rm rel}$ the relative velocity, and the elastic scattering cross section $\sigel$
from nucleons, which in our case only include protons.

\begin{figure}
\includegraphics[width=0.22\textwidth]
{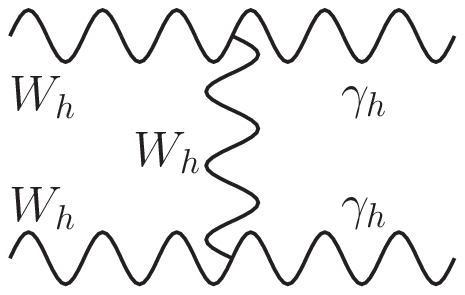}
\includegraphics[width=0.22\textwidth]
{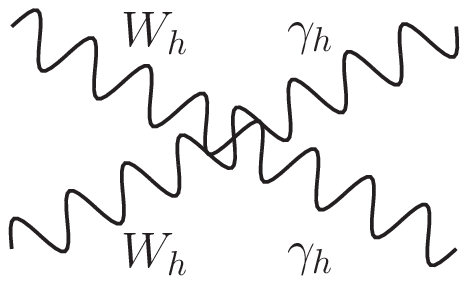}
\caption{Leading annihilation process within the set of simplifying assumptions in the text.
There is also a third annihilation diagram that is obtained from the left one by a crossing.}
\label{ann}
\end{figure}
For simplicity of our treatment, while maintaining the key aspects of the model, we will henceforth
assume $\mah \ll \mwh$ with $\eta\equiv v_\phi/v_\Phi \ll 1$, and ${\mzh, \,M_\Phi > 2 \mwh}$, where $M_\Phi$ is the mass of $\Phi$.
Then, the dominant process that would set the relic density of $W_h$ is $W_h W_h \to
\gamma_h \gamma_h$, given by the Feynman diagrams in Fig.~\ref{ann}.  The $s$-channel
annihilation through $\Phi$ is suppressed by the small $\Phi\gamma_h \gamma_h$ coupling
which is proportional to $\eta^4$.  This suppression does not apply to $\phi \gamma_h \gamma_h$ coupling
and one has to consider the effect of $\Phi$-$\phi$ mixing in the $s$-channel diagram.  The scalar
mixing will come from a term $\lambda_m \phi^\dagger \phi \Phi^\dagger \Phi$.  For
a perturbative self-coupling of $\phi$, we then expect $\mu_\phi \lsim v_\phi$, where
$\mu_\phi$ is the mass parameter in the $\phi$ potential.
Hence, unless $\mu_\phi$ is set by a tuned cancellation, we
must assume $\lambda_m v_\Phi^2 \lsim v_\phi^2$ which yields $\lambda_m \lsim \eta^2$.  Thus,
$\Phi$-$\phi$ mixing is typically suppressed by $\eta^2$.

These processes are governed
by the $W_hW_h \gamma_h$ and $W_h W_h \gamma_h \gamma_h$ vertices whose Lorentz structure is
identical to the familiar analogues in the SM.
However, the overall coupling here is set by $g_h \sh$.
We find that the thermally averaged annihilation cross section is well approximated by
\beq
\vev{\sigannv} \simeq \frac{19\, (g_h \sh)^4}{72 \pi \mwh^2}\,,
\label{sigannv}
\eeq
where the mass of $\gamma_h$ has been ignored.  More detailed results,
including the $p$-wave contributions, are given in the Appendix and
are used to obtain our numerical results presented below.
The relic density of $W_h$ is given by \cite{KolbTurner}
\beq
\Omega_h h^2\simeq 1.04 \times 10^9 \frac{x_f\, {\rm GeV}^{-1}}{\sqrt{g_\star}\, M_{\rm Pl}\, \vev{\sigannv}}\,,
\label{Omegah}
\eeq
where $g_\star$ is the number of relativistic degrees of freedom at the time of freeze-out
and $M_{\rm Pl} \simeq 1.22 \times 10^{19}$~GeV is the Planck mass.  The quantity
$x_f = \mwh/T_f$, with $T_f$ the freeze-out temperature, is given by \cite{KolbTurner}
\beq
x_f \simeq \ln [0.038 (\kappa/\sqrt{x_f g_\star}) M_{\rm Pl} \mwh \vev{\sigannv}]\,,
\label{xf}
\eeq
where $\kappa=3$ for a massive vector boson.  The insertion of $\vev{\sigannv}$ in Eqs.~(\ref{Omegah}) and (\ref{xf})  are valid for only the $s$-wave approximation.  A more complete expression for the relic density, including $p$-wave contributions, is given in the Appendix.  For the range of parameters relevant in
our work $x_f\simeq 20$ is a good approximation and we will use this value in
the following.

If dark matter
is composed only of $W_h$, then the relic density $\Omega_h h^2\simeq 0.12$~\cite{Ade:2013zuv} can be used to solve for $g_h\sh$ in terms of $M_{W_h}$.  Under this assumption and ${\mah\ll \mwh}$, we find
\beq
(g_h \sh)^2 \simeq \frac{M_{W_h}}{10~{\rm GeV}}\left\{\begin{array}{l l} 2.2\times 10^{-3};\quad & T_f\lesssim\Lambda_{QCD}\\ 1.5\times 10^{-3}; & T_f\gtrsim\Lambda_{QCD} \end{array}\right.,
\label{gh2}
\eeq
where the $p$-wave expansion and $x_f\simeq 20$ has been used.  The two solutions are due to the different counting of degrees of freedom below and above the QCD phase transition, which we assume to occur at $\Lambda_{QCD}\simeq 200~{\rm MeV}$.  For freeze-out temperatures below the QCD phase transition we use $g_\star = 13.75$ accounting for the neutrinos, electron, photon, and $\gamma_h$; for freeze-out temperatures above the QCD phase transition we additionally include the muon, gluons, and $u,d,s$ quarks in the counting and find $g_\star = 64.75$.  While the inclusion of $\gamma_h$ in $g_\star$ depends critically on $\mah$ and charm quarks (muons) should be included for $T_f\gtrsim M_c$ ($\Lambda_{QCD}\gtrsim T_f\gtrsim M_\mu$), we find that these considerations only make $\sim 5\%$ corrections in our determination of $(g_h\sh)^2$.  Hence, for simplicity we neglect these effects in the numerical results that follow.

An implicit assumption in the above derivation was that dark matter initially starts
in thermal equilibrium before freeze-out .
For consistency, we would then require that the
hidden photon $\gamma_h$ decay rate $\Gamma_{\gamma_h}$
into SM final states is large enough to keep
up with the expansion of the Universe at $T=T_f$~\cite{KolbTurner}. Assuming $\mah\lesssim 1$~GeV, the decay rate is
\begin{eqnarray}
\Gamma_{\gah}&\simeq&\frac{\alpha}{3}\left(\eps \cos\theta_h\right)^2\mah\sum_{F}N_{C}Q^2_F\left(1+\frac{2M_F^2}{M^2_\gah}\right)\beta_F\nonumber\\
&\lesssim&\frac{4\alpha}{3}\left(\eps \cos\theta_h\right)^2\mah,
\end{eqnarray}
where  $\alpha \equiv e^2/(4\pi) \simeq 1/137$, $\beta_F=\sqrt{1-4M_F^2/\mah^2}$,
and the sum is over fermions, $F$, with masses ${2\,M_F\le \mah}$, charges $Q_F$, and colors $N_C=1$ for leptons and $N_C=3$ for quarks.
The inequality is obtained for ${F\in \{e, \mu, u, d, s\}}$.  The expansion rate at freeze-out
is set by the Hubble constant $H(T_f)= 1.7 g_{\star}^{1/2} T_f^2/M_{\rm Pl}$.  For $\mah<T_f$ the requirement that $\gah$ are in thermal equilibrium at $T_f$ is satisfied when~\cite{KolbTurner}
\beq
\frac{\mah}{T_f}\Gamma_{\gamma_h}\gsim H(T_f)\,.
\label{thermeq}
\eeq 
Using this condistion and $T_f=\mwh/20$ yields
\beq
(\eps\, \cos\theta_h)^2\left(\frac{\mah}{\rm MeV}\right)^2 \gsim 10^{-12}g^{1/2}_{\star}\left(\frac{\mwh}{10~{\rm GeV}}\right)^3,
\label{eps2m}
\eeq
which does not give an important limitation on $\eps\,\cos\theta_h$ for scenarios of interest in our work,
as we will see later. For $\mah\gtrsim$~MeV, this also means that the $\gamma_h$ lifetime
is much shorter than the Hubble time $\sim 1$~s associated with
Big Bang Nucleosynthesis and $\gamma_h$
decays will not affect primordial nuclear processes.

The scalars in our model are expected to decay promptly. For example, given the assumption
${M_\Phi>2 \mwh}$ above, $\Phi$ decays will be prompt, since they occur at tree-level and
are not kinematically suppressed.  The same applies to $\phi$ decays as long as $M_\phi\gsim \mah$,
since even for one off-shell $\gamma_h$, we expect a rate of order $\eps^2 g_h'^2 \alpha$ and
for values of $\eps$ that would be of interest here these decays will be prompt compared to the
relevant cosmological time scales.
\section{Direct Detection}

\begin{figure}
\includegraphics[width=0.48\textwidth]
{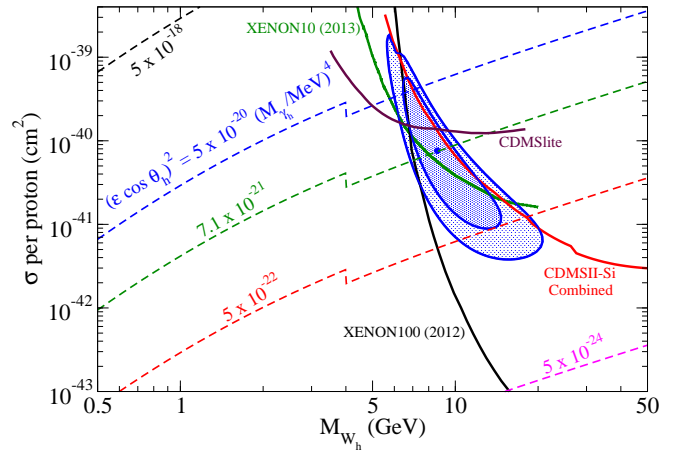}
\caption{The elastic cross section for various values of $\eps \cos\theta_h$ as a function of
$\mwh$.  Constraints form direct detection experiments are also included.  The dot corresponds
to the CDMSII-Si highest likelihood point \cite{Agnese:2013rvf}.  The inner contour surrounding the dot is the 68\% C.L. and the outer the 90\% C.L. }
\label{sigmap}
\end{figure}

Another important quantity for a dark matter candidate is its direct detection
cross section.  In the case of our $W_h$ candidate, this is governed by its scattering
from protons in the nucleus by exchanging $Z_h$ or $\gamma_h$.  Under our assumption $M_{\gamma_h}\ll M_{Z_h}$, the $\gamma_h$ exchange is the dominant contribution to the scattering process.  The elastic
scattering cross section from a nucleus $N$, with atomic number $Z$,
is then given by
\beq
\sigel \simeq \frac{4\, Z^2\,\alpha \,(\eps\,\cos\theta_h)^2\,(g_h\,\sin\theta_h)^2 \, \mu_{\rm r}^2(W_h, N)}{ M_\gah^4}\,,
\label{sigel}
\eeq
where  $\mu_{\rm r}(X, Y) = M_X\, M_Y /(M_X + M_Y)$ is the reduced mass of the system;
$M_N = A \, m_n$, $A$ is the mass number, and $m_n \simeq 938$~MeV is the mass of a nucleon $n$.
In the above, terms higher order in dark matter velocity are ignored
and nuclear form factors have been set to unity (a good simplifying approximation).
The elastic cross section per {\it proton} is then given by \cite{sigmap}
\beq
\sigma_p \simeq
\frac{4\,\alpha \,(\eps\,\cos\theta_h)^2 \,(g_h\,\sin \theta_h)^2 \, \mu_{\rm r}^2(W_h, n)}{\mah^4}\,.
\label{sigman}
\eeq
The elastic cross section per {\it nucleon} is obtained from $\sigma_n = (Z^2/A^2) \sigma_p$.

In Fig.~\ref{sigmap}, we have plotted $\sigma_p$ versus $\mwh$,
assuming $W_h$ is dark matter for various values of $\eps \cos \theta_h$ as
a function of $\mah$.
For consistency, we have chosen values of parameters that would
yield the correct relic abundance from Eq.~(\ref{Omegah})
for each value of $\mwh$.  In this plot, we have also presented
various relevant constraints on $\sigma_p$ at 90\% confidence level (C.L.) from XENON10~\cite{Angle:2011th}, XENON100~\cite{Aprile:2012nq}, the combined CDMSII-Si data~\cite{Agnese:2013rvf,Agnese:2013cvt}, and CDMSlite~\cite{Agnese:2013lua}.  The dot in
the plot marks maximum likelihood point (8.6~GeV, $\sigma_n=1.9\times 10^{-41}$~cm$^2$, $\sigma_p=7.6\times 10^{-41}$~cm$^2$)
from the CDMSII-Si data \cite{Agnese:2013rvf}.  We note that
XENON10 and XENON100 data mildly disfavor this signal.
However, there may be considerations that could lead to
loosened constraints on this point in the parameter space \cite{Hooper:2013cwa}.

It is amusing to note that a lower bound on $\mah$ can be obtained by combining an observation of DM at a direct detection experiment and the requirement that $\gah$ remain in thermal equilibrium until $T_f$.  Once $\sigma_p$ and $\mwh$ are measured, Eq.~(\ref{sigman}) can be combined with the relic density constraint Eq.~(\ref{gh2}) to obtain a relation between $\eps\sin\theta_h$ and $M_\gah$.  This relation can then be used in conjunction with Eq.~(\ref{eps2m}) to obtain a lower bound on $\mah$:
\begin{eqnarray}
\frac{\mah}{40~{\rm MeV}}&\gtrsim& \left(\frac{\mwh}{10~\rm{GeV}}\right)^{2/3}\left(\frac{\mu_r(W_h,n)}{1~\rm{GeV}}\right)^{1/3}\label{mahbnd}\\
&&\times\left(\frac{\sigma_p}{8\times10^{-41}~{\rm cm}^2}\right)^{-1/6}\,.\nonumber
\end{eqnarray}
This lower bound was derived using the condition in Eq.~(\ref{thermeq}) which is valid for $\mah<T_f$~\cite{KolbTurner}. For ${\mwh\gtrsim 1}$~GeV and $\sigma_p\gtrsim 10^{-43}$~cm$^2$, the $\mah$ lower bound given above is below $T_f$ and remains valid.

\section{Fixed Target and Dark Matter Beam Experiments}
\begin{figure}[t]
\centering
\includegraphics[width=0.48\textwidth,clip]{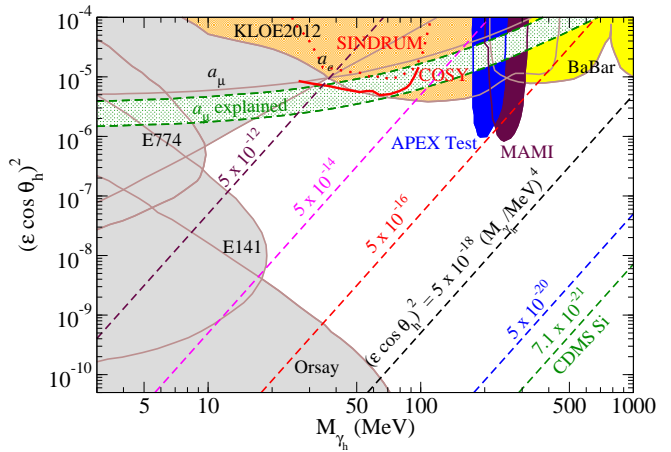}
\caption{Shaded regions indicate parameter space ruled out by various experiments.  The shaded green region indicates parameter region that accommodate the muon $g_\mu-2$ anomaly.  The line $(\varepsilon\cos\theta_h)^2=7.1\times 10^{-21}(M_{\gamma_h}/{\rm MeV})^4$ is consistent with the CDMSII-Si anomaly.}
\label{current}
\end{figure}

There has been much interest in searching for new light weakly coupled vector bosons in  fixed target and beam dump experiments~\cite{Bjorken:2009mm,Reece:2009un,Andreas:2012mt,Merkel:2011ze,Bossi:2009uw,Abrahamyan:2011gv,Gninenko:2012eq}.   In these experiments, when an electron scatters off the target it bremsstrahlungs a light gauge boson, which subsequently decays to a lepton pair.  New light vector bosons can also be searched for in low energy $e^+e^-$ experiments~\cite{Reece:2009un,Essig:2009nc} and in meson decays~\cite{Bjorken:2009mm,Reece:2009un,Babusci:2012cr,MeijerDrees:1992kd,Gninenko:2013sr,Adlarson:2013eza}.  Such searches typically depend on finding resonances in the resulting dilepton spectrum.

\begin{figure}[t]
\centering
\includegraphics[width=0.48\textwidth,clip]{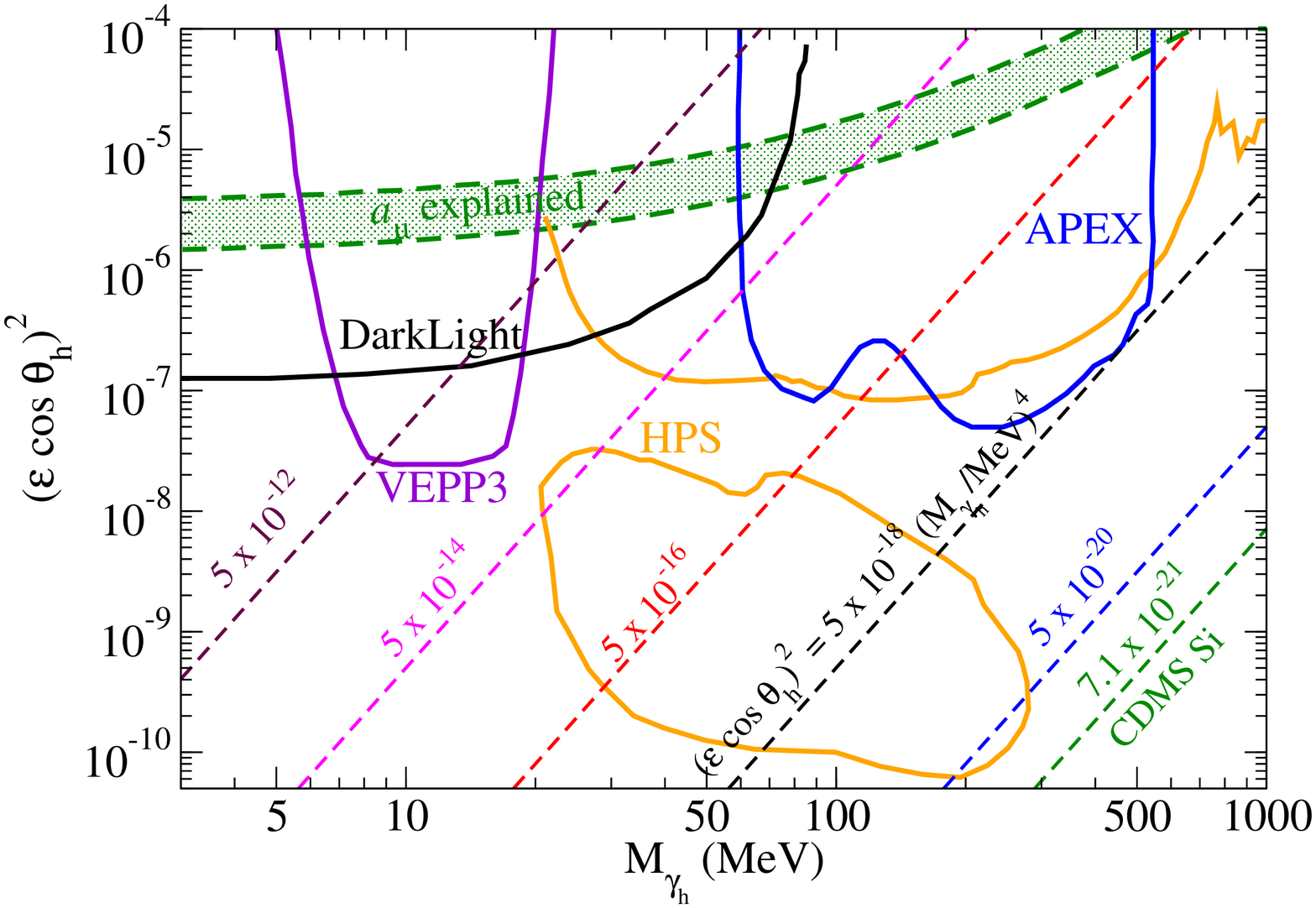}
\caption{Future sensitivities at fixed target experiments.  The line corresponding to $(\varepsilon\cos\theta_h)^2=7.1\times 10^{-21}(M_{\gamma_h}/{\rm MeV})^4$ is consistent with the CDMSII-Si anomaly.}
\label{future}
\end{figure}

The model presented here contains two vector bosons, $\gamma_h$ and $Z_h$, that couple to the SM electromagnetic current via kinetic mixing as shown in Eq.~(\ref{EMcouplings}).  Under our assumptions that $M_{Z_h}\geq 2M_{W_h}$ we expect $Z_h\rightarrow W_h W_h$ to be the dominant decay channel since the $Z_h$ couplings to the SM are suppressed compared to the $Z_h-W_h$ coupling. However, with our assumptions, the only decay channels available for $\gamma_h$ is into light SM fermions.  Hence, only $\gamma_h$ is expected to contribute to dilepton signals significantly.

In Fig.~\ref{current} we show the current bounds from low energy $e^+e^-$ experiments~\cite{Reece:2009un,Essig:2009nc}, meson decays ($\pi^0$ at WASA-at-COSY~\cite{Adlarson:2013eza} and SINDRUM~\cite{MeijerDrees:1992kd,Gninenko:2013sr}, $\phi$ at KLOE~\cite{Babusci:2012cr}, and $\Upsilon$ at BaBar~\cite{Bjorken:2009mm,Reece:2009un}), fixed target experiments (MAMI~\cite{Merkel:2011ze} and APEX~\cite{Abrahamyan:2011gv}), and the electron dipole moment~\cite{Davoudiasl:2012ig,Endo:2012hp}.  The green shaded region indicates the parameter region consistent with the muon magnetic moment, $g_\mu-2$, anomaly~\cite{Fayet:2007ua,Pospelov:2008zw,Davoudiasl:2012qa,Bennett:2006fi}.  We include dashed lines of parameter combinations that are relevant for the direct detection of $W_h$.  The sensitivities to the $\varepsilon\cos\theta_h$ and $M_{\gamma_h}$ are to be compared to the direct detection experiment sensitivities in Fig.~\ref{sigmap}.

In Fig.~\ref{future} we show the sensitivity of the future fixed target experiments HPS, DarkLight, APEX, and VEPP3~\cite{Wojtsekhowski:2012zq}.  For an overview of the HPS, DarkLight, and APEX experiments please see Ref.~\cite{Boyce:2012ym}.  We note that the VEPP3 experiment is insensitive to the decay products of the light vector boson.  Comparing to Fig.~\ref{sigmap}, it can be seen that these future experiments begin to be sensitive to $\varepsilon\cos\theta_h$ and $M_{\gamma_h}$ combinations that are relevant to light DM direct searches.

There are also proposals to measure invisibly decaying hidden bosons at fixed target experiments~\cite{DMbeam1,DMbeam2}.  As before, hidden bosons are produced during interactions between the incident beam and the target.  Subsequently, the hidden boson decays produce a beam of DM which then scatters off the nuclei of a detector, similar to direct detection experiments (however, here DM is relativistic).
Typically it is assumed that the same hidden boson produces the DM and mediates
the scattering off nuclei.  However, in the scenario presented here, the dark matter would be produced via $Z_h$ and then
recoil dominantly via $\gamma_h$, since this is a $t$-channel process.  We may hence expect the  scenario presented here to
manifest itself in many distinct experimental scenarios with tightly correlated parameter spaces.

\section{Conclusions}
In this work, we considered a hidden sector gauge symmetry $\Gh$, broken
completely by a Higgs doublet of $SU(2)_h$ and a Higgs singlet field
charged only under $U(1)_h$.  Our assumed gauge sector provides a simple
setup to illustrate possible effects of non-abelian interactions that may
govern the dynamics of DM.  The hidden sector was only assumed to
interact with the SM indirectly, through kinetic mixing of $U(1)_h$ and
hypercharge $U(1)_Y$.
This setup was shown to yield a good DM candidate, corresponding
to hidden non-abelian gauge fields, removing the need to assume
extra fermions or scalars for DM.  The stability of DM is guaranteed
by a $Z_2$ remnant of the original gauge symmetry.  Hence, in this
framework DM is a manifestation of hidden or dark forces of Nature. Our model
can readily lead to the correct relic density and an allowed direct detection
cross section.  In particular, the model can accommodate the recent tentative
signal from CDMSII-Si \cite{Agnese:2013rvf} corresponding to DM
masses $\sim 10$~GeV.

Since our framework has an expanded hidden gauge sector, compared
to models where only a dark $U(1)$ is assumed, we find that the hidden and
the SM sectors interact via exchange of two distinct vectors bosons.  These vector bosons
are coupled to the electromagnetic current of the SM with a suppressed coupling governed
by the degree of kinetic mixing.  Typically, one would then expect signals
corresponding to two different resonances that are often
called ``dark photons" in the $U(1)$ models.
Therefore, for light DM, the well-studied
phenomenology of dark photons at low energies can be relevant to our model.

We also note that the two dark photons that arise in our model can be naturally
hierarchic in mass and the heavier vector can in principle decay to vector DM.  This
process is mediated by an analogue of the SM $Z W^+ W^-$ vertex, where DM
is the hidden counter part to $W^\pm$.  In this case, the heavy
vector is mainly invisible, since it would dominantly decay into DM, given its
suppressed dark-photon-like coupling to the SM sector.  Hence, our model can
lead to ``dark matter beam" signals in fixed target experiments that have been
proposed in the context of $U(1)$ models. In the typical DM beam scenario, the same vector that produces the DM is also responsible for DM scattering in the detector.  In our setup, the DM is produced via the decay of the heavy vector while the detection is dominated by the exchange of the light vector.  Thus, here the production and detection of the DM beam are mediated by two different vectors, the
lighter of which is expected to have mainly visible decay modes.

We note that
the light vectors of the hidden dynamics in our model may contribute to
novel rare Higgs decay signals~\cite{Davoudiasl:2013aya}. Also, present day annihilation of $W_h$ in the galactic halo may lead to potential GeV scale indirect detection signals. The totality of the signals available
in this framework can in principle be used to discern the hidden gauge
sector and its pattern of symmetry breaking, and potentially
lead to the conclusion that ``Dark Force = Dark Matter."

\acknowledgments

Work supported in part by the United States Department of
Energy under Grant Contracts DE-AC02-98CH10886.


\appendix
\makeatletter
\renewcommand\section{\@startsection {section}{1}{\z@}%
    {-3.5ex \@plus -1ex \@minus -.2ex}%
    {2.3ex \@plus.2ex}%
    {\normalfont\large\bfseries}}
\makeatother
\renewcommand{\thesubsection}{\Alph{section}.\arabic{subsection}}
\noindent
\section{Mass Diagonalization}
\subsection{Hidden Sector}
The dark sector consists of two ``charged" gauge bosons $W^{\pm}_h$, and two ``neutral" gauge bosons $W^3_h,B_h$.  The masses of $W^{\pm}_h$ are then given simply by
\beq
\mwh = \frac{g_h}{2}\,v_\Phi.
\label{mwh}
\eeq
In the $(B_h,W^3_h)$ basis, the ``neutral" mass matrix is
\beq
\frac{1}{4}\begin{pmatrix}
  {g_h'}^2(v_\Phi^2+v_\phi^2) & -g_h g_h' v_\Phi^2 \\
  -g_h g_h' v_\Phi^2      & g_h^2 v_\Phi^2
   \end{pmatrix}.
\label{2x2orig}
\eeq
The above matrix is then diagonalized via the transformation
\beq
\begin{pmatrix}B_h\vspace{0.02in}\\ W^3_h\end{pmatrix}=\begin{pmatrix*}[r]\cos\theta_h & -\sin\theta_h\vspace{0.02in}\\\sin\theta_h & \cos\theta_h\end{pmatrix*}\begin{pmatrix} \gamma^0_h\vspace{0.02in}\\Z^0_h\end{pmatrix},
\label{HidDiag}
\eeq
where $\gamma^0_h$ and $Z^0_h$ are mass eigenstates of the $2\times2$ matrix with masses $M_{\gamma^0_h}$ and $M_{Z^0_h}$, respectively, such that ${M_{Z^0_h}> M_{\gamma^0_h}}$.

In analogy with the SM weak mixing angle, we introduce a new angle completely determined by the hidden sector gauge couplings
\beq
\tan\zeta=g_h'/g_h.
\eeq
The mass matrix in Eq.~(\ref{2x2orig}) can then be written in the somewhat simpler form
\beq
\mwh^2
\begin{pmatrix}
(1+\eta^2)~\tan^2\zeta & -\tan\zeta\\
-\tan\zeta & 1
\end{pmatrix},
\label{2x2}
\eeq
where, again, $\eta=v_\phi/v_\Phi$ is the ratio of the hidden sector Higgs vevs.
The masses for $\gamma^0_h$ and $Z^0_h$ can then be expressed as
\begin{eqnarray}
M^2_{Z^0_h}+M^2_{\gamma^0_h}&=&\frac{\mwh^2}{\cos^2\zeta}\left(1+\eta^2\sin^2\zeta \right)\\
\frac{M^2_{Z^0_h}-M^2_{\gamma^0_h}}{M^2_{Z^0_h}+M^2_{\gamma^0_h}}&=&\sqrt{1-\frac{\eta^2\sin^22\zeta}{(1+ \eta^2\sin^2\zeta)^2}}.
\end{eqnarray}
The mixing angle $\theta_h$ can be solved for and we get
\beq
\cos^2 \theta_h = \frac{\mwh^2 - M_{\gamma^0_h}^2}{M_{Z^0_h}^2-M_{\gamma^0_h}^2}\,.
\eeq
Since we have already assumed $M_{Z^0_h}\geq M_{\gamma^0_h}$, the positivity of $\cos^2\theta_h$ enforces the hierarchy $M_{W_h}\geq M_{\gamma^0_h}$.  Similarly, the positivity of $\sin^2\theta_h$ can be used to show $M_{Z^0_h}\geq\mwh$.  We then naturally have the mass hierarchy $M_{Z^0_h}\geq \mwh\geq M_{\gamma^0_h}$.

An interesting limit to look at is $\eta\ll 1$.  From Eq.~(\ref{2x2}), it is clear that the effect of the singlet Higgs $\phi$ is a mild perturbation in the usual SM neutral vector boson mass matrix.  Explicitly, we find
\begin{eqnarray}
M_{\gamma^0_h}&=& M_{W_h} \eta \sin\zeta+\mathcal{O}(\eta^3)\\
M_{Z^0_h}&=&\frac{\mwh}{\cos\zeta}\left(1+\frac{1}{2}\eta^2\sin^4\zeta\right)+\mathcal{O}(\eta^4)\\
\tan\theta_h&=&\tan\zeta\left(1+\eta^2\sin^2\zeta\right)+\mathcal{O}(\eta^4).
\end{eqnarray}
As expected, this is the same as the SM result plus small perturbations in $\eta$.  In the limit $\eta\rightarrow 0$ the SM
mass relations are exactly reproduced.

\subsection{Kinetic Mixing}

Now we consider the effects of kinetic mixing:
\beq
\mathcal{L}_{\rm kin}=-\frac{1}{4}\hat{B}_{\mu\nu}\hat{B}^{\mu\nu}+\frac{1}{2}\frac{\varepsilon}{\cos\theta_W}\hat{B}_{\mu\nu}\hat{B}_h^{\mu\nu}-\frac{1}{4}\hat{B}_{h\mu\nu}\hat{B}^{\mu\nu}_h\,.
\eeq
The kinetic terms is diagonalized via the shifts
\begin{eqnarray}
\hat{B}^\mu&=&B^\mu+\frac{\varepsilon/\cos\theta_W}{\sqrt{1-\varepsilon^2/\cos^2\theta_W}}B_h^\mu\\
\hat{B}_h^\mu&=&\frac{1}{\sqrt{1-\varepsilon^2/\cos^2\theta_W}}B_h^\mu,
\end{eqnarray}
where $B^\mu$ is now identified as the SM hypercharge and $B_h$ is identified as the $U(1)_h$ gauge boson.

The transformation of $\hat{B}_h$ does not involve $B$ and so does not introduce any SM fields into the hidden sector covariant derivative.  However, the transformation of $\hat{B}$ does introduce $B_h$ into the SM covariant derivative.  Hence, all induced mass mixing comes from the SM Higgs vev and not the hidden sector.  Specifically, after diagonalizing the kinetic term, the SM covariant derivative is
\begin{eqnarray}
D^{SM}_\mu&=&\partial_\mu+i\left(\frac{g'\xi}{\sin\theta_W}YB_{h\mu}\right.\\
&&\left.+g'YB_\mu+gT^3W^3_\mu+\frac{g}{\sqrt{2}}(\tau^+W^+_\mu+\tau^-W^-_\mu)\right)\,,\nonumber
\end{eqnarray}
where $Y$ is the SM hypercharge operator, $T^3$ is the SM isospin operator, $\tau^\pm$ are the charged current operators, and $g$ and $g'$ are the
$SU(2)$ and $U(1)_Y$ SM gauge couplings, respectively. The parameter
\beq
\xi = \varepsilon\tan\theta_W/\sqrt{1-\varepsilon^2/\cos^2\theta_W}
\eeq
 has been introduced to simplify notation.  Performing the usual SM rotation ${B_\mu=\cos\theta_W A_\mu-\sin\theta_W \hat{Z}^0_\mu}$ and ${W^3_\mu=\sin\theta_W A_\mu +\cos\theta_W \hat{Z}^0_\mu}$; as well as the rotation in Eq.~(\ref{HidDiag}) the SM covariant derivative is
\begin{eqnarray}
D^{SM}_\mu&=&\partial_\mu+i\left(\frac{g'\xi}{\sin\theta_W}Y(\cos\theta_h \gamma^0_{h\mu}-\sin\theta_h Z^0_{h\mu})\right.\\
&&\left.+eQA_\mu+g_Z Q_Z \hat{Z}^0_\mu+\frac{g}{\sqrt{2}}(\tau^+W^+_\mu+\tau^-W^-_\mu)\right)\,,\nonumber
\end{eqnarray}
where $e$ and $Q=T^3+Y$ are the usual electromagnetic charge and operator; $g_Z=e/(\sin\theta_W\cos\theta_W)$ and $Q_Z=T^3\cos^2\theta_W-Y\sin^2\theta_W$ are the SM neutral current coupling and operator, respectively.

It is now clear that once the Higgs vev is inserted into the Higgs kinetic term, there is mass mixing between the $\hat{Z}^0$, $\gamma^0_h$, and $Z^0_h$ proportional to the kinetic mixing parameter $\varepsilon$.  In the $(\hat{Z}^0,\gamma^0_h,Z^0_h)$ basis, the mass matrix is
\beq
M^2_{Z^0}\begin{pmatrix}
1 & -\xi \cos\theta_h & \xi\sin\theta_h \\
-\xi\cos\theta_h & \mu^2_{\gamma^0_h}+\xi^2\cos^2\theta_h & -\xi^2\sin\theta_h\cos\theta_h \\
\xi\sin\theta_h & -\xi^2\sin\theta_h\cos\theta_h & \mu^2_{Z^0_h}+\xi^2\sin^2\theta_h
\end{pmatrix},
\eeq
where $M_{Z^0}=g_Z v/2$ is the SM $Z$ mass and ${\mu_i = M_i/M_{Z^0}}$.  To leading order in $\varepsilon$, the $3\times3$ mass matrix can be diagonalized with the transformations
\begin{eqnarray}
\gamma^0_{h\mu}&\simeq&\gamma_{h\mu}-\frac{\varepsilon\tan\theta_w\cos\theta_h}{1-\mu^2_{\gamma^0_h}} Z^0_\mu\label{Ahfin}\\
Z^0_{h\mu}&\simeq&Z_{h\mu}+\frac{\varepsilon\tan\theta_W\sin\theta_h}{1-\mu^2_{Z^0_h}} Z^0_\mu\label{Zhfin}\\
\hat{Z}^0_\mu&\simeq& Z^0_\mu+\\
&&\varepsilon\tan\theta_W\left(\frac{\cos\theta_h}{1-\mu^2_{\gamma^0_h}}\gamma_{h\mu}-\frac{\sin\theta_h}{1-\mu^2_{Z^0_h}}Z_{h\mu}\right)\nonumber
\label{Zfin},
\end{eqnarray}
where $\gamma_h,\,Z_h$ and $Z^0$ are the mass eigenstates with masses
\begin{eqnarray}
M_{\gamma_h}^2&\simeq&M^2_{\gamma^0_h}\left(1-\xi^2 \frac{ \cos^2\theta_h}{1-\mu^2_{\gamma^0_h}}\right)\label{ahmass}\\
M_{Z_h}^2&\simeq& M^2_{Z^0_h}\left(1-\xi^2\frac{\sin^2\theta_h}{1-\mu^2_{Z^0_h}}\right)\label{zhmass}\\
M^2_Z&\simeq&M^2_{Z^0}\times\label{Zmass}\\
&&\left(1+\xi^2\frac{1-\mu^2_{Z^0_h}\cos^2\theta_h-\mu^2_{\gamma^0_h}\sin^2\theta_h}{(1-\mu^2_{\gamma^0_h})(1-\mu^2_{Z^0_h})}\right),\nonumber
\end{eqnarray}
respectively. We have kept to $\mathcal{O}(\varepsilon^2)$ since this is the lowest order of $\varepsilon$ in Eqs.~(\ref{ahmass}-\ref{Zmass}).  Using these masses we see that Eq.~(\ref{cos2thetah}) is valid to $\mathcal{O}(\varepsilon^2)$.

Finally, the interactions between the gauge bosons with fermions and scalars are governed by the covariant derivative.  Applying the final transformation to the mass eigenstate, to $\mathcal{O}(\varepsilon^2)$ the SM covariant derivative is
\begin{eqnarray}
D^{SM}_{\mu}&=&\partial_\mu+i\frac{g}{\sqrt{2}}(\tau^+W^+_\mu+\tau^-W^-_\mu)\\
&&+ie Q A_\mu+ig_Z Q_Z Z_\mu\nonumber\\
&&+i\varepsilon e\cos\theta_h\left(Q+Q_Z\sec^2\theta_W\frac{\mu^2_{\gamma^0_h}}{1-\mu^2_{\gamma^0_h}}\right)\gamma_{h\mu}\nonumber\\
&&-i\varepsilon e\sin\theta_h\left(Q+Q_Z\sec^2\theta_W\frac{\mu^2_{Z^0_h}}{1-\mu^2_{Z^0_h}}\right)Z_{h\mu}\,.\nonumber
\end{eqnarray}
The interactions between $\gamma_h~(Z_h)$ and the SM neutral current is suppressed by an additional factor of $\mu^2_{\gamma^0_h}~ (\mu^2_{Z^0_h})$.

We also note that the transformations in Eqs.~(\ref{Ahfin},\ref{Zhfin}) will induce couplings between the SM $Z^0$ and the dark sector currents.  However, these currents are suppressed by $\varepsilon$ and not expected to make significant contribution to the electroweak precision observables.

\section{Relic Density Calculation}
For the reference of the reader, we briefly review the calculation of relic density and obtain a general formula.
\subsection{Thermally averaged cross section}
The calculation of the thermally averaged cross section follows Ref.~\cite{Srednicki:1988ce}.  The general  annihilation process
\beq
\chi(p_1)+\chi(p_2)\rightarrow X(p_X)
\eeq
is considered, where $\chi$ is a DM candidate and $X$ is some final state that may be multi-particle.  The usual annihilation cross section is then
\beq
\sigma_{\rm ann} v_{\rm rel}=\frac{1}{4 E_1 E_2} \int\overline{|\mathcal{M}(\chi\chi\rightarrow X)|}^2 dPS_{X}\,,
\eeq
where $v_{\rm rel}$ is the relative speed of the DM particles, $\overline{|\mathcal{M}|}^2$ is the spin summed and averaged matrix element squared, and $dPS_{X}$ is the final state phase space.

The thermally averaged cross section is evaluated by performing a weighted integral over the possible initial state momentum configurations
\beq
\vev{\sigannv} \equiv \frac{\kappa^2}{n^2_{eq}}\int \frac{d^3p_1}{(2\pi)^3} \frac{d^3p_2}{(2\pi)^3} f(E_1)f(E_2) \sigannv,
\label{thermave}
\eeq
where $f(E)$ is the energy distribution of particle $\chi$, $\kappa$ is the number of internal degrees of freedom of $\chi$, and $n_{eq}=\kappa/(2\pi)^3 \int d^3p f(E)$ is the equilibrium density.

In general, $\sigannv$ is dependent upon the reference frame in which it is evaluated.  Following Ref.~\cite{Srednicki:1988ce}, it is useful to introduce the Lorentz invariant quantity
\beq
w(s) = \frac{1}{4}\int\overline{|\mathcal{M}(\chi\chi\rightarrow X)|}^2 dPS_{X}.
\eeq
After integration over final state phase space, the function $w$ can only depend on the particle masses and $s=(p_1+p_2)^2$.  Hence, $w$ can be evaluated in a specific frame and then generalized to an arbitrary frame with the identification:
\beq
s=2(M^2_\chi+E_1E_2-p_1p_2\cos\theta_{12}),
\eeq
where $\theta_{12}$ is the angle between the initial state momenta.

We are interested in temperatures such that $x=M_\chi/T\gg 1$ and will express the final result as an expansion in $1/x$.  In this limit, the energy distributions are well approximated by the Boltzmann distribution $f(E)=\exp(-E/T)$. The integrals in Eq.~(\ref{thermave}) are simplified by the change of variables
\beq
y_i=\frac{E_i-M_\chi}{T}=\frac{E_i}{T}-x,
\eeq
for $i=1,2$.  Using these variables, we can solve for
\begin{eqnarray}
z\equiv\frac{s}{4M_\chi^2}&=&1+\frac{1}{2x}(y_1+y_2)+\frac{1}{2x^2}y_1y_2\\
&&-\frac{1}{x}\sqrt{\left(y_1+\frac{1}{2x}y^2_1\right)\left(y_2+\frac{1}{2x}y^2_2\right)}\cos\theta_{12}\,.\nonumber
\end{eqnarray}
Hence, an expansion in $1/x$ can be obtained by expanding $w(s)$ around $s=4M_\chi^2$ and the rest of the integrand around $1/x$.  The final result is
\begin{eqnarray}
\vev{\sigannv}=\frac{1}{M_\chi^2}\left.\left[w-\frac{3}{2x}(2w-w')+\mathcal{O}(x^{-2})\right]\right|_{z=1}\!\!\!,~~~
\label{thermavefin}
\end{eqnarray}
where $w'=\partial w/\partial z$.  The $\mathcal{O}(x^0)$ term is typically referred to as the $s$-wave and the $\mathcal{O}(x^{-1})$ as the $p$-wave.

For completeness, we provide the full result of the thermally averaged cross section for $W_h\,W_h\rightarrow \gah\,\gah$, as shown in Fig.~\ref{ann}, up to the $p$-wave expansion:
\begin{widetext}
\begin{eqnarray}
\vev{\sigannv}&=&\frac{19 (g_h\sin\theta_h)^4}{72\pi \mwh^2}(1-r_{\gamma_h}^2)^{-1/2}\left(1-\frac{r_{\gamma_h}^2}{2}\right)^{-4}\\
&&\times\left[\left(1-\frac{55}{19}r^2_{\gah}+\frac{295}{76}r^4_{\gah}-\frac{241}{76}r^6_{\gah}+\frac{233}{152}r^8_{\gah}-\frac{121}{304}r^{10}_{\gah}+\frac{33}{608}r^{12}_{\gah}-\frac{3}{608}r^{14}_{\gah}\right)\right.\nonumber\\
&&\left.-\frac{39}{19 x}\left(1-\frac{433}{156}r^2_{\gah}+\frac{349}{78}r^4_{\gah}-\frac{2851}{624}r^6_{\gah}+\frac{7}{3}r^8_{\gah}-\frac{5}{8}r^{10}_{\gah}+\frac{77}{832}r^{12}_{\gah}-\frac{41}{4992}r^{14}_{\gah}\right)\right],\nonumber
\end{eqnarray}

\end{widetext}
where $r_{\gah}=M_{\gah}/\mwh$.

\subsubsection{Velocity Expansion}
The above result has been explicitly constructed as an expansion of the thermal integral and $w$ in the variables $x=M_{\chi}/T$ and $z=s/4M_{\chi}^2$.  Often one expands of the annihilation cross section in terms of the relative velocity
\beq
\sigannv = a+b\, v^2_{\rm rel}\,.
\eeq
Using the relation for the relative velocity
\beq
v^2_{\rm rel}=|{\bf v_1}-{\bf v_2}|^2 = v^2_1\,+\,v^2_2\,-2\,v_1\,v_2\,\cos\theta_{12}\,,
\label{vrel}
\eeq
the thermally averaged cross section is then
\beq
\vev{\sigannv}=a+\frac{6b}{x}\,,
\label{thermaveNorm}
\eeq
where $\bf{v_{1,2}}$ are the velocities of the initial state particles, and we have used the thermodynamic relation $\vev{v^2_{\rm rel}}=6/x$.

The advantage of using Eq.~(\ref{thermavefin}) to calculate the thermally averaged cross section is that $w$ is Lorentz invariant and is expanded in terms of a Lorentz invariant quantity. Hence, there are no frame-dependent ambiguities in calculating $\vev{\sigannv}$. However, obtaining the usual result in Eq.~(\ref{thermaveNorm}) from Eq.~(\ref{thermavefin}) is not completely transparent.  To illustrate how this is accomplished and that the two are equivalent, we now {\it derive} Eq.~(\ref{thermaveNorm}) from Eq.~(\ref{thermavefin}).

First rewrite $w'$ in Eq.~(\ref{thermavefin}) as a derivative with respect to $v^2_{\rm rel}$.  Since $w$ is Lorentz invariant, this operation can performed in the center of momentum frame:
\beq
v^2_{\rm rel} = 4\,v^2_1=4\,v^2_2=4\left(1-\frac{1}{z}\right)\,,
\eeq
where $v_i$ is the magnitude of the vector $\bf{v_{i}}$. Hence, evaluating $w$ and its derivatives at $v_{\rm rel}=0$ is equivalent to $z=1$.

For ease of notation, the velocity expansion of $w$ is expressed as
\beq
w(s)=\alpha+\beta\,v^2_{\rm rel},
\eeq
where $\alpha$ and $\beta$ are constants.
The thermally averaged cross section is then
\beq
\vev{\sigannv}=\frac{1}{M_{\chi}^2}\left[\alpha-\frac{3}{x}(\alpha-2\beta)+\mathcal{O}(x^{-2})\right]
\eeq

To obtain Eq.~(\ref{thermaveNorm}), $\alpha$ and $\beta$ need to be determined in terms of $a,b$.  Since $\sigannv$ is not Lorentz invariant, we perform this in an arbitrary frame.  That is we use the relative velocity in Eq.~(\ref{vrel})
and the relation
\beq
w(s)=E_1 E_2 \sigannv\,.
\eeq
Expanding both sides to $\mathcal{O}(v^2_1,v^2_2,v_1v_2)$ and integrating over $\theta_{12}$, we find
\beq
\alpha=M_{\chi}^2\,a,\quad{\rm and}\quad \beta=M_{\chi}^2\left(b+\frac{a}{2}\right).
\eeq
Using these results and Eq.~(\ref{thermavefin}), we obtain the well-known result in Eq.~(\ref{thermaveNorm}).

\subsection{Relic Density}
The relic density of a particle $\chi$ is typically given in terms of the variable $\Omega = \rho(0)/\rho_{\rm crit}$, where $\rho(0)$ is the present energy density of particle $\chi$.  The critical density is $\rho_{\rm crit}=3H^2(0)/8\pi G$ with $G$ being Newton's constant, $H(T)=\dot{R}/R$ the expansion rate of the universe, and $R$ the cosmic scale factor.  Since the relic $\chi$ is massive, at present day $x\gg1$ and the energy density is $\rho(0) = M_\chi n(0)$, where $n(T)$ is the number density at temperature $T$.  Hence, the number density needs to be solved for.  The details of this derivation are well-known and can, for example, be found in Refs.~\cite{KolbTurner,Srednicki:1988ce}.

The number density, $n$, of particle $\chi$ obeys the Boltzmann equation
\beq
\frac{d n}{dt}=-3\,H(T)\,n-\vev{\sigannv}\,(n^2-n^2_{eq})\,.
\eeq
Using well-known methods~\cite{KolbTurner,Srednicki:1988ce}, this differential equation can be simplified to
\beq
\frac{dY}{dx}=-\frac{\vev{\sigannv}\,s(T)}{x\,H(T)}(Y^2-Y^2_{eq})\,,
\eeq
where $Y=n/s$ and $s$ is the entropy density.

To evaluate $n(0)$, late times $x\gg 1$ need to be considered.  After freeze-out, the number density $n$ is stable while the equilibrium number density $n_{eq}$ continues to decrease as the photon temperature continues to drop.  Hence we have $Y\gg Y_{eq}$ and the differential equation simplifies to
\beq
\frac{dY}{dx}\simeq-\frac{\vev{\sigannv}s(T)}{x H(T)}Y^2\,.
\label{Ydiff}
\eeq
To solve this equation we expand the thermally averaged cross section
\beq
\vev{\sigannv}=\sum_j a_j x^{-j}.
\eeq
Furthermore, the effective degrees of freedom are introduced
\begin{eqnarray}
g_{\star s}(T)=\frac{45}{2\pi^2}\frac{s(T)}{T^3},\quad
g_{\star}(T)=\frac{30}{\pi^2}\frac{\rho(T)}{T^4},
\end{eqnarray}
and the solution for $H$ in Robertson-Walker metric is used:
\beq
H(T)=\left(\frac{8\pi}{3}G\,\rho(T)\right)^{1/2}.
\eeq

Equation~(\ref{Ydiff}) is then integrated from freeze out, ${x=x_f}$, to present-day, ${x=+\infty}$, to find
\beq
Y_\infty =\left(\frac{45}{\pi}\right)^{1/2}\frac{1}{M_{\rm Pl}M_{\chi}}\frac{1}{g^{-1/2}_\star g_{\star s}}\frac{x_f}{\sum_j a_j x_f^{-j}/(j+1)},
\eeq
where $Y_\infty$ is the present-day value and $g_\star$ and $g_{\star s}$ are evaluated at freeze-out.

Finally, putting everything together, the relic density is
\beq
\Omega h^2=\frac{5.36\times 10^{43}\,{\rm cm}^3\, {\rm GeV}\, s(0)}{M^3_{\rm Pl}\,g_\star^{-1/2}g_{\star s}}\frac{x_f}{\sum_j a_j x_f^{-j}/(j+1)},
\eeq
where $H(0)=(100\, {\rm km}~{\rm s}^{-1}~{\rm Mpc})~h$ and $G=1/M_{\rm Pl}^2$ have been used, and the present day entropy density is $s(0)=2889.2$~cm$^{-3}$~\cite{Beringer:1900zz}.  To a good approximation, at freeze-out $g_\star\simeq g_{\star S}$ and we use this in the numerical solutions presented in the text.

\end{document}